\documentclass[12pt]{iopart}
\usepackage{amssymb,graphicx,color}

\newcommand{\eqref}[1]{(\ref{#1})}

\newcommand{\eps}{\varepsilon}

\newcommand{\lam}{\lambda}

\newcommand{\si}{\sigma}

\newcommand{\pa}{\partial}

\newcommand{\dlam}{\dot{\lambda}}
\newcommand{\la}{\langle}
\newcommand{\ra}{\rangle}

\newcommand{\nn}{\nonumber}
\newcommand{\hL}{\hat{L}_\lambda\,}
\newcommand{\hLa}{\hat{L}^+_\lambda\,}
\newcommand{\baf}{\bar{f}}
\newcommand{\bW}{\bar{W}}

\begin{document}

\title[On the work distribution in quasi-static processes]{On the work distribution in quasi-static processes}

\author{Johannes Hoppenau}
\address{Universit\"at Oldenburg, Institut f\"ur Physik, 26111 Oldenburg,
  Germany}
\author{Andreas Engel}
\address{Universit\"at Oldenburg, Institut f\"ur Physik, 26111 Oldenburg,
  Germany}

\begin{abstract}
We derive a systematic, multiple time-scale perturbation expansion for the work distribution in isothermal quasi-static Langevin processes. To first order we find a Gaussian distribution reproducing the result of Speck and Seifert [Phys. Rev. E {\bf 70}, 066112 (2004)]. Scrutinizing the applicability of perturbation theory we then show that, irrespective of time-scale separation, the expansion breaks down when applied to untypical work values from the tails of the distribution. We thus reconcile the result of Speck and Seifert with apparently conflicting exact expressions for the asymptotics of work distributions in special systems and with an intuitive argument building on the central limit theorem.
\end{abstract}

\maketitle

\section{Introduction}
Thermodynamics of systems so small that typical changes of their energies are of the order of $k_B T$ is concerned with {\em distributions} of thermodynamic quantities rather than with their averages \cite{Udorev,Jarrev,Esprev}. These distributions fulfill a number of exact and general relations which are now commonly referred to as fluctuation theorems. At least two features of these theorems are truly remarkable: First, they hold for systems driven (almost) arbitrarily far from equilibrium, and second, they are very sensitive to the {\em tails} of the respective probability distributions \cite{hugorev}.

The concept of work is a cornerstone of thermodynamics. Accordingly, the distribution of work is of central importance in stochastic thermodynamics. If the parameters of a system coupled to a heat bath are changed very slowly one expects that the system stays in equilibrium during the whole process. In this case the work performed or consumed in the transition is not fluctuating and equals the difference in free energy between final and initial state of the system. If, on the other hand, the system is driven violently it will pass through intermediate non-equilibrium states. In this case a fluctuating fraction of {\em dissipative} work adds to the free-energy difference. The resulting distribution of work is, in general, non-universal displaying features specific for the system and the process at hand.

In an attempt to establish some universality even for non-equilibrium processes Speck and Seifert investigated the work distribution for isothermal stochastic processes with small but {\em non-zero} driving \cite{SpSe}. By using a projection-operator technique to derive approximate solutions of the underlying Fokker-Planck equation \cite{Risken} they found that in these cases the work distribution must be Gaussian to leading order. A simple and intuitive argument put forward in \cite{HeJa} relates this result to the central limit theorem. In a quasi-static process the system has time to relax to its instantaneous equilibrium distribution between changes of the external parameters. Therefore, the total work becomes a sum of many {\em independent} contributions and by virtue of the central limit theorem the resulting distribution has to converge to a Gaussian. 

On the other hand, the determination of the exact {\em asymptotics} of the work distribution for a simple model system found an exponential tail for {\em any driving} \cite{NiEn}. Hence, the Gaussian character of the work distribution of quasi-static processes does not extend to the rare realizations. This is again in accordance with the central limit theorem which prescribes a Gaussian for the {\em central} part of the distribution leaving the tails unspecified \cite{Feller}. From the systematic analysis in \cite{SpSe} it appeared, however, that time scale separation alone is sufficient to derive a Gaussian form of the work distribution.

In the present note we analyze the work distribution in quasi-static stochastic processes by employing a multiple-scale perturbation expansion. The expansion parameter quantifies the time-scale separation in the problem. We first rederive the results of Speck and Seifert and then show that, {\em irrespective of time-scale separation}, the expansion fails for the  tails of the work distribution. A similar argument applies to the projection operator technique used by Speck and Seifert. We illustrate the point with numerical simulations for a model system and also indicate how the expansion may be extended to higher orders.

\section{The model}
As a special yet representative example we consider the overdamped dynamics of a degree of freedom $x$ in a time-dependent potential $V(x,\lam)$. The protocol $\lam(t)$ specifies the time-dependence of the potential and evolves from $\lam(0)=0$ at the beginning of the process to $\lam(t_f)=1$ at its end. Upon redefinition of $\lam$ we may always achieve a linear protocol, $\lam=t/t_f$. For notational simplicity we restrict ourselves to scalar $x$, generalization to higher dimensions being straightforward. The dynamics is given by
\begin{equation}\label{LEx}
 \pa_t x(t)=-\mu \, \pa_x V(x,\lam)+\zeta(t)\; ,
\end{equation}  
where $\mu$ denotes the mobility and $\zeta$ is a standard white-noise source with correlation 
\begin{equation}
 \la \zeta(t) \zeta(t')\ra=\frac{2\mu}{\beta}\,\delta(t-t')\; .
\end{equation} 
Here $\beta$ denotes the inverse temperature of the heat bath. The work $W$ performed along a trajectory $x(t)$ is given by \cite{Jar97,Udorev}
\begin{equation}\label{defW}
 W[x(\cdot)]=\int_0^{t_f}\!\! dt\; \dlam\, \pa_\lam V(x(t),\lam(t))\; ,
\end{equation}  
hence
\begin{equation}\label{LEW}
 \pa_t W=\dlam \, \pa_\lam V(x,\lam)\; ,
\end{equation} 
where the dot denotes the time derivative.

The joint probability distribution $p(x,W,t)$ describing the coupled stochastic evolution of $x$ and $W$ as given by \eqref{LEx} and \eqref{LEW} therefore fulfills the Fokker-Planck equation 
\begin{equation}\label{FPE}
 \pa_t p= \mu\pa_x\Big((\pa_x V) p\Big)-\dlam\,  \pa_\lam V \,\pa_W p +\frac{\mu}{\beta}\,\pa_x^2\, p\; .
\end{equation} 
It is convenient to write this equation in the form 
\begin{equation}\label{FPE2}
 \pa_t p= (\hL+\hat{L}^W_\lam)\, p
\end{equation} 
with 
\begin{equation}
 \hL:=\mu\pa_x\Big(\pa_x V(x,\lam) +\frac{1}{\beta}\,\pa_x\Big)
\end{equation} 
and 
\begin{equation}
 \hat{L}^W_\lam:=-\dlam\, \pa_\lam V(x,\lam)\, \pa_W\; .
\end{equation} 

For each {\em fixed} value $\lam$ of the protocol parameter $\hL$ has a right eigenvector with eigenvalue zero given by the equilibrium distribution 
\begin{equation}\label{deff}
 f_\lam(x):=e^{\beta(F_\lam-V(x,\lam))}
\end{equation} 
with the free energy 
\begin{equation}\label{defF}
 F_\lam:=-\frac{1}{\beta}\ln \int dx\; e^{-\beta V(x,\lam)}\; .
\end{equation}
The corresponding left eigenvector is given by $\baf(x)\equiv 1$. We assume that the process starts in equilibrium. According to \eqref{defW} the initial condition for \eqref{FPE} therefore reads
\begin{equation}\label{ic}
 p(x,W,0)=f_0(x)\delta(W)\; .
\end{equation} 
Our aim is to characterize the marginal work distribution
\begin{equation}\label{defP}
 P(W,t):=\int dx \; p(x,W,t)
\end{equation}   
in the quasi-static limit, i.e. when $\lam$ changes sufficiently slowly. 


\section{Multiple time-scale perturbation theory}
The concept of a quasi-static process rests on {\em time-scale separation}. For fixed $\lam$ the relaxation of the marginal distribution $f(x,t):=\int dW \, p(x,W,t)$ to its equilibrium form, $f_\lam(x)$, happens with a characteristic time $\tau_\mathrm{rel}$, which, in general, depends on $\lam$. We call a process {\em quasi-static}, if  
\begin{equation}\label{defqust}
 \dlam(t)\,\tau_\mathrm{rel}\ll 1 \qquad \forall t\in(0,t_f)\; .
\end{equation} 
Physically, this means that $\lam(t)$ hardly changes over times of order $\tau_\mathrm{rel}$ or, equivalently, that on the time scale on which $\lam$ evolves $f(x,t)$ is always very near to $f_{\lam(t)}(x)$. Choosing $\tau_\mathrm{rel}$ as time unit it is therefore natural to use $\eps:=\dlam=1/t_f$ as a small parameter and to investigate the problem perturbatively.

For $\eps=0$ no work at all is performed and $P(W,t)=\delta(W)$ for all $t$. The form of $P(W,t)$ changes qualitatively when $\eps$ becomes non-zero, i.e. the perturbation is {\em singular} and a multiple-scale perturbation scheme is called for \cite{KeCo,BeOr}. Accordingly, we introduce an additional {\em slow} time variable $T:=\eps t=\lam$ and perform the substitutions
\begin{equation}
 p(x,W,t) \rightarrow p(x,W,t,\lam),\qquad 
 \pa_t \rightarrow \pa_t+\eps\pa_\lam\; .
\end{equation}  

Moreover, since our focus is on the work distribution evolving on the slow time scale $\lam$, we are not interested in the fast dynamics and neglect the dependence of $p(x,W,t,\lam)$ on $t$ altogether. The Fokker-Planck equation \eqref{FPE2} then acquires the form
\begin{equation}\label{FPEpert}
 \hL p(x,W,\lam)=\eps\, \big(\pa_\lam+\pa_\lam V(x,\lam)\, \pa_W\big)\, p(x,W,\lam)
\end{equation} 
appropriate for a perturbation expansion. The singular character of the perturbation is apparent from the fact that the time derivative is among the small terms. 

We solve \eqref{FPEpert} iteratively using the ansatz
\begin{equation} \label{pertansatz}
 p(x,W,\lam)=p_0(x,W,\lam)+\eps  p_1(x,W,\lam)+ \eps^2  p_2(x,W,\lam) +\dots\; .
\end{equation}  
Omitting the $t$-dependence of $p$ may be interpreted as part of this ansatz.


Plugging \eqref{pertansatz} into \eqref{FPEpert} and matching powers of $\eps$ we generate a hierarchy of equations for the unknown functions $p_n(x,W,\lam)$. To zeroth order we find 
\begin{equation}
 \hL p_0=0
\end{equation}  
with the solution 
\begin{equation}\label{resp_0}
 p_0(x,W,\lam)=f_\lam(x) g_0(W,\lam)\; ,
\end{equation} 
where $g_0(W,\lam)$ is a so far undetermined function constrained only by the normalization condition   
\begin{equation}
 \int dW g_0(W,\lam)=1\qquad \forall \lam\; .
\end{equation} 

To order $\eps$ we get 
\begin{equation}\label{eqp_1}
 \hL p_1=(\pa_\lam+\pa_\lam V\, \pa_W)\, p_0
        =(\pa_\lam f_\lam)\, g_0 + f_\lam(\pa_\lam g_0+\pa_\lam V\, \pa_W g_0)\; .
\end{equation} 
In order for the perturbation expansion to be well-defined and secular terms being absent the r.h.s. of this equation must be orthogonal to the null space of the adjoint operator
\begin{equation}\label{defLa}
 \hLa=-\mu (\pa_x V(x,\lam))\,\pa_x+\frac{\mu}{\beta}\,\pa_x^2
\end{equation} 
of $\hL$, i.e. orthogonal to $\baf(x)\equiv 1$. We hence impose the {\em solvability condition}
\begin{equation}\label{solv1}
 0=\int dx \Big[(\pa_\lam f_\lam) g_0 + f_\lam (\pa_\lam g_0+\pa_\lam V\, \pa_W g_0)\Big]\; .
\end{equation}
Now 
\begin{equation}\label{normf}
 \int dx \, \pa_\lam f_\lam(x)=\pa_\lam \int dx f_\lam(x)=0
\end{equation}
and \eqref{solv1} translates into
\begin{equation}\label{resg_0}
 \pa_\lam g_0=-A_0\pa_W g_0
\end{equation} 
with 
\begin{equation}\label{defA_0}
 A_0(\lam):=\la \pa_\lam V\ra_\lam:=\int dx\, f_\lam(x)\, \pa_\lam V(x,\lam) 
           =\pa_\lam F_\lam\; .
\end{equation} 
Using \eqref{resg_0} as well as 
\begin{equation}
 \pa_\lam f_\lam=\beta(A_0-\pa_\lam V)f_\lam
\end{equation} 
as follows from \eqref{deff} in the r.h.s. of \eqref{eqp_1} we get 
\begin{equation}\label{eq2p_1}
 \hL p_1(x,W,\lam)=f_\lam(x) a_0(x,\lam)\Big[\beta g_0(W,\lam)-\pa_W g_0(W,\lam)\Big]\; ,
\end{equation} 
where
\begin{equation}
 a_0(x,\lam):=A_0(\lam) -\pa_\lam V(x,\lam)\; .
\end{equation} 
Eq. \eqref{eq2p_1} may now be solved. Observing
\begin{equation}
 \hat{L}_\lam^{-1} (f_\lam(x) h(x))=f_\lam(x) (\hLa)^{-1} h(x)
\end{equation} 
for any function $h(x)$ \cite{Risken} we may write the solution in the form
\begin{equation}\label{resp_1}
 p_1(x,W,\lam)=f_\lam(x) b_0(x,\lam)\Big[\beta g_0(W,\lam)-\pa_W g_0(W,\lam)\Big]
               +f_\lam(x) g_1(W,\lam)
\end{equation} 
where 
\begin{equation}\label{defb_0}
 b_0:=(\hLa)^{-1} a_0
\end{equation} 
denotes the normalizable solution of the equation $\hLa b_0=a_0$.
The second term on the r.h.s. of \eqref{resp_1} is the general solution of the homogeneous equation $\hat{L}_\lam p_1 = 0$. By a redefinition of $g_1$ we may include any contribution proportional $f_\lam(x)$ from the first term into this second one. Without loss of generality we may therefore require
\begin{equation}\label{normb_0}
 \int dx \baf(x) f_\lam(x) b_0(x,\lam)=\la b_0\ra_\lam=0\; .
\end{equation}

At order $\eps^2$ we have 
\begin{equation}\label{eqp_2}
 \hL p_2=(\pa_\lam+\pa_\lam V \pa_W) p_1\; .
\end{equation} 
Using \eqref{resp_1}, \eqref{normf}, and \eqref{normb_0} the solvability condition for this equation takes the form
\begin{equation}\label{resg_1}
 \pa_\lam g_1=-A_0\pa_W g_1 -A_1(\beta \pa_W g_0-\pa^2_W g_0)\; ,
\end{equation}  
with 
\begin{equation}\label{defA_1}
 A_1(\lam):=\la \pa_\lam V\,b_0\ra_\lam\; .
\end{equation} 
These results are sufficient to derive an equation for $P(W,\lam)$ correct up to order $\eps$. To this order it is not necessary to actually solve \eqref{eqp_2}.


\section{The Gaussian approximation}
From \eqref{defP}, \eqref{resp_0}, \eqref{resp_1}, and \eqref{normb_0} we find 
\begin{equation}\label{exp1P}
 P(W,\lam)=g_0(W,\lam)+\eps g_1(W,\lam) + {\cal O}(\eps^2)\; .
\end{equation} 
Therefore
\begin{equation}
 \pa_\lam P=\pa_\lam g_0 +\eps \pa_\lam g_1 + {\cal O}(\eps^2)
\end{equation} 
which using \eqref{resg_0} and \eqref{resg_1} gives rise to
\begin{eqnarray}\nn
  \pa_\lam P&=& -A_0\pa_W g_0-\eps A_0\pa_W g_1 -\eps A_1(\beta \pa_W g_0-\pa^2_W g_0)
               + {\cal O}(\eps^2)\\\nn
   &=& -A_0 \pa_W (g_0+\eps g_1)- \eps A_1(\beta \pa_W g_0-\pa^2_W g_0)+ {\cal O}(\eps^2)\\\nn
   &=& -A_0 \pa_W P-\eps A_1(\beta \pa_W P-\pa^2_W P)+ {\cal O}(\eps^2)\\
   &=& -(A_0+\eps \beta A_1)\pa_W P+\eps A_1 \pa_W^2 P+{\cal O}(\eps^2)\; .\label{eq1P}
\end{eqnarray}
The solution of this equation satisfying the initial condition \eqref{ic} is a Gaussian
\begin{equation}\label{res1P}
 P(W,\lam)=\frac{1}{\sqrt{2\pi\si_W^2(\lam)}}
             \exp{\left(-\frac{(W-\bW(\lam))^2}{2\si_W^2(\lam)}\right)}
\end{equation} 
with 
\begin{equation}\label{res1av}
 \si_W^2(\lam)=2\eps \int_0^\lam\, d\lam' A_1(\lam')
\end{equation} 
and 
\begin{equation}\label{res1var}
 \bW(\lam)=\int_0^\lam d\lam'\, [A_0(\lam')+\eps \beta A_1(\lam')]
          =\Delta F(\lam) +\frac{\beta}{2} \si_W^2(\lam)\; .
\end{equation} 
Eqs. \eqref{res1P}--\eqref{res1var}  reproduce the results of Speck and Seifert \cite{SpSe}. 


\section{Self-consistency of the perturbation expansion}
We next address the question under which conditions our perturbative treatment is self-consistent. To this end it is instructive to consider the distribution in $x$  conditioned to a given value of $W$. From \eqref{pertansatz}, \eqref{exp1P}, as well as from  \eqref{resp_0} and \eqref{resp_1} we find
\begin{eqnarray}\label{defpcond}
 p(x,\lam|W)&:=&\frac{p(x,W,\lam)}{P(W,\lam)}\nn\\
    & =& f_\lam(x)\frac{g_0+\eps g_1+\eps b_0(\beta g_0-\pa_W g_0)+{\cal O}(\eps^2)}
                      {g_0+\eps g_1 + {\cal O}(\eps^2)}\nn\\
    & =& f_\lam(x)\Big[1+\eps b_0(\beta-\pa_W \ln P)+{\cal O}(\eps^2)\Big]\; .
\end{eqnarray}
For perturbation theory to be applicable we hence need 
\begin{equation}
 \eps |b_0(x,\lam)(\beta- \pa_W \ln P(W,\lam))|\ll 1\; .
\end{equation} 
Consequently, $\eps\ll 1$, as ensured by time-scale separation alone, is {\em not sufficient} to make the perturbation expansion meaningful. In the present context we are not concerned with the singular character of the small-noise limit, $\beta\to\infty$, and assume $\beta={\cal O}(1)$. The crucial criterion for the self-consistency of our perturbation expansion is therefore
\begin{equation}\label{criterion}
 \eps |b_0 \, \pa_W \ln P|\ll 1\; .
\end{equation} 
From \eqref{res1P} we find $|\pa_W \ln P(W)|=|W-\bW|/\si^2_W$. Hence \eqref{res1P} gives a reliable approximation of the true work distribution for {\em typical values} of $W$ only. For values of $W$ sufficiently different from its average (as measured in units of the standard deviation $\si_W$) $P(W,\lam)$ may significantly differ from the Gaussian form. In fact, since $b_0$ does not depend on $W$ there is for every value of $\eps$ a threshold $W_c$ such that \eqref{res1P} substantially deviates from the true work distribution when $|W|\gg W_c$. 

\begin{figure}[t]
  \centering
  \includegraphics[width=.5\columnwidth]{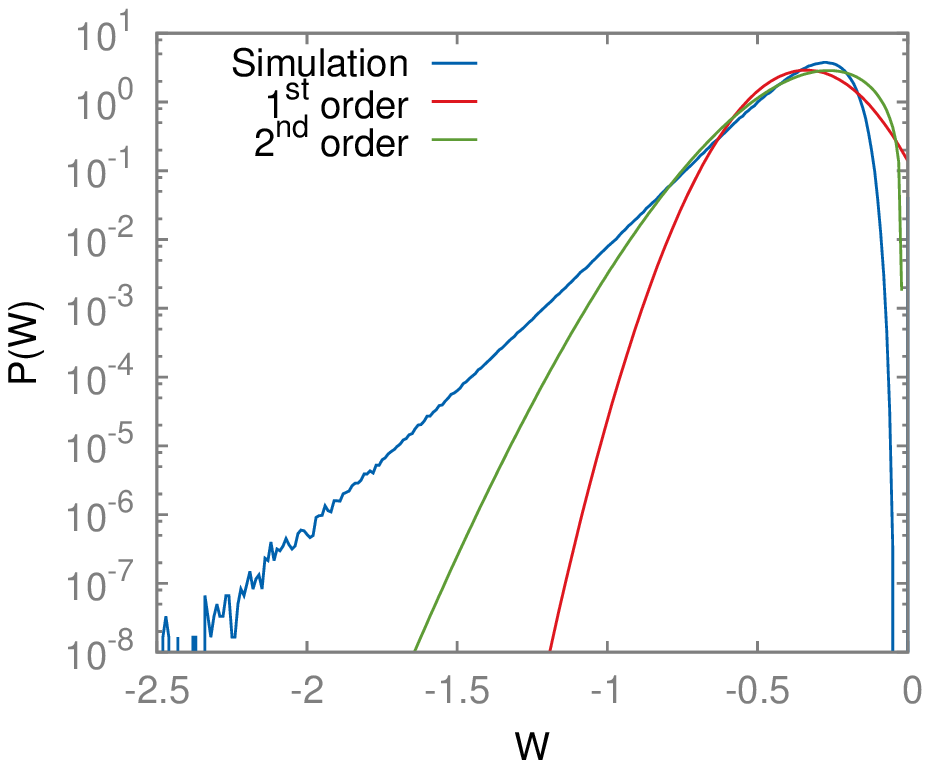}%
  \includegraphics[width=.5\columnwidth]{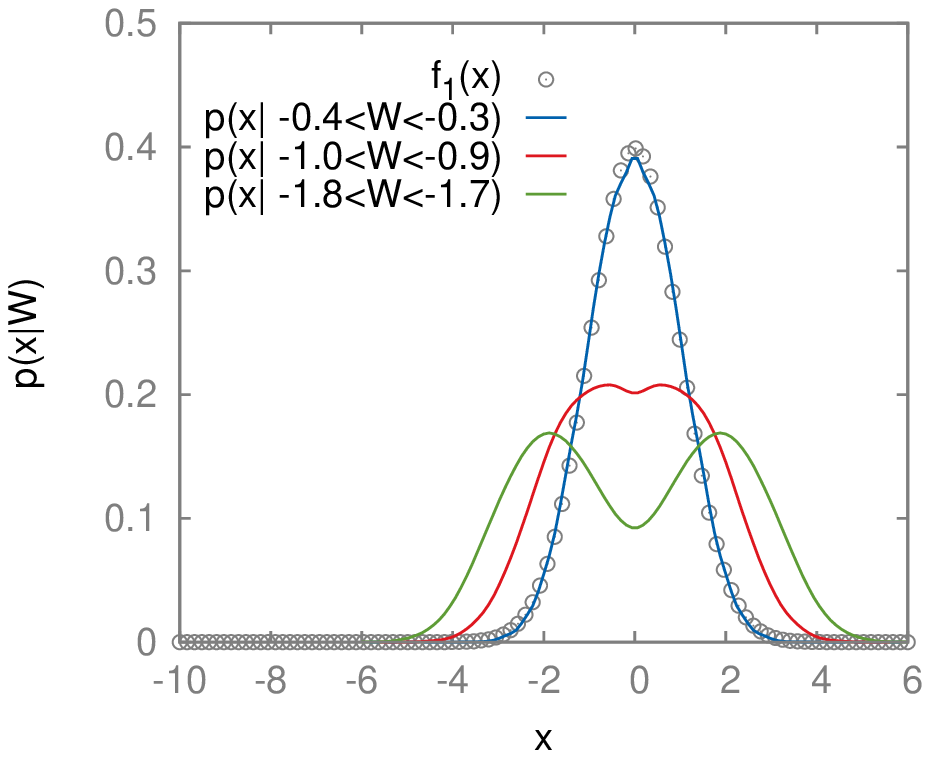}
  \caption{Left: Results from a numerical simulation of Eq.~\eqref{LEx} for the ``breathing parabola'' \eqref{defbreathpar} with $\mu=\beta=1$ and $t_f=10$ together with the perturbative results of first and second order. Note the strong deviation between the distribution $P(W)$ as obtained from the simulations (blue line) and the Gaussian approximation \eqref{res1P} (red line). Right: Comparison of the final equilibrium distribution $f_1(x)$ with the conditional distribution $p(x|W)$ for several intervals of $W$. All histograms are compiled from the simulation of $6\cdot10^9$ trajectories, the conditional distributions are smoothed by a spline-interpolation.}
  \label{fig1}
\end{figure}

From \eqref{defpcond} we realize that the failure of perturbation theory is related to the fact that $p(x,\lam|W)$ may be rather different from $f_\lam(x)$. In fact this makes perfect sense. Typical values of $W$ are produced by typical trajectories $x(t)$ the endpoints  $x(t_f)$ of which provide a fair sampling of $f_\lam(x)$. Conditioning the distribution of $x$ to {\em unusual} work values, on the other hand, implies a strong bias on the weights of the trajectories which generically result in substantial deviations of $p(x,\lam|W)$ from the equilibrium distribution $f_{\lam}(x)$. To illustrate this fact we show in Fig.~\ref{fig1} results obtained by simulating \eqref{LEx} for the ``breathing parabola''  
\begin{equation}\label{defbreathpar}
 V(x,\lam)=\frac{2-\lam}{2} \, x^2
\end{equation} 
with $\mu=\beta=1$ and $t_f=10$. The left figure displays the work distribution on a logarithmic scale and shows that for most $W$-values the distribution is indeed not Gaussian. In the right figure we compare the equilibrium distribution $f_\lam(x)$ with $p(x,\lam|W)$ for different intervals of $W$-values. As expected, choosing $W$ from the center of $P(W,\lam)$ gives rise to a $p(x,\lam|W)$ rather similar to $f_\lam(x)$. Very small values of $W$, however, require trajectories $x(t)$ with sufficiently large excursions from $x=0$ such that $p(x,\lam|W)$ broadens and finally develops a bimodal shape that is even qualitatively different from the Gaussian equilibrium distribution. Note that for the potential chosen $\tau_\mathrm{rel}=1/(\mu(2-\lam))\leq 1$ for the entire process. Together with $\dlam=1/t_f=0.1$ time-scale separation as defined by \eqref{defqust} is hence well satisfied. 


\section{Higher orders}
It is straightforward to extend our perturbation expansion to higher orders. Enforcing \eqref{resg_1} we rewrite \eqref{eqp_2} in the form 
\begin{eqnarray}\nn
 \hL p_2&=&f_\lam\Big[a_0(\beta g_1-\pa_W g_1)
                +(\beta a_0 b_0+\pa_\lam b_0)(\beta g_0-\pa_W g_0)\\\label{eq2p_2}
      & & \qquad\qquad -(A_0 b_0 +a_1)(\beta \pa_W g_0-\pa^2_W g_0)\Big],
\end{eqnarray} 
with
\begin{equation}
 a_1(x,\lam):=A_1(\lam)-\pa_\lam V(x,\lam) b_0(x,\lam)\; .
\end{equation} 
The solution of \eqref{eq2p_2} is given by  
\begin{equation}\label{resp_2}
 p_2=f_\lam\Big[b_0(\beta g_1-\pa_W g_1)+b_1(\beta g_0-\pa_W g_0)
     -c_1(\beta \pa_W g_0-\pa^2_W g_0)+g_2\Big],
\end{equation} 
where we have introduced the normalizable solutions 
\begin{eqnarray}
 b_1(x,\lam)&:=&(\hLa)^{-1}(\beta a_0 b_0+\pa_\lam b_0)\quad\mathrm{and}\\
 c_1(x,\lam)&:=&(\hLa)^{-1}(A_0 b_0+a_1)\; .
\end{eqnarray}
as well as another solution $f_\lam(x)g_2(W,\lam)$ of the homogeneous equation. 
Similarly to \eqref{normb_0} we may assume that
\begin{equation}\label{normb_1c_1}
 0=\la b_1\ra_\lam (\beta g_0-\pa_\lam g_0)-\la c_1\ra_\lam (\beta \pa_W g_0-\pa^2_W g_0)\; .
\end{equation}
Using \eqref{resp_2} we may explicitly calculate the r.h.s. of the order $\eps^3$ equation
\begin{equation}
 \hL p_3=(\pa_\lam-\pa_\lam V \pa_W) p_2\; .
\end{equation}  
The solvability condition at this order takes the form
\begin{eqnarray}\nonumber
 \pa_\lam g_2&=&-A_0\pa_W g_2 - A_1(\beta \pa_W g_1-\pa^2_W g_1)
              - A_2(\beta \pa_W g_0-\pa^2_W g_0)\\
             & &\qquad\qquad +C_2(\beta \pa^2_W g_0-\pa^3_W g_0),
\end{eqnarray} 
with 
\begin{equation}
 A_2(\lam):=\la \pa_\lam V b_1 \ra_\lam \qquad\mathrm{and}\qquad
 C_2(\lam):=\la \pa_\lam V c_1 \ra_\lam \; .
\end{equation}
Proceeding in the same way as we did in deriving \eqref{eq1P} we end up with  
\begin{eqnarray}\nonumber
 \pa_\lam P &=& -(A_0+\eps\beta A_1+\eps^2\beta A_2) \pa_W P
              +(\eps A_1 +\eps^2 (A_2+\beta C_2))\pa_W^2 P\\\label{eq2P}
            & &\qquad\qquad  -\eps^2 C_2 \pa^3_W P  + {\cal O}(\eps^3)\; .
\end{eqnarray} 
The last term describes deviations from a Gaussian $P(W,\lam)$ which hence first show up at order $\eps^2$. 

As a simple example we have carried through this program for the potential \eqref{defbreathpar}. The results are 
\begin{equation}
 A_0=-\frac{1}{2\beta(2-\lam)},\quad A_1=\frac{1}{4\mu\beta^2(2-\lam)^3} \; ,
\end{equation} 
and 
\begin{equation}
 A_2=-\frac{3}{8\mu^2\beta^2(2-\lam)^5},\quad C_2=-\frac{1}{4\mu^2\beta^3(2-\lam)^5}\; .
\end{equation} 
The left plot in Fig.~\ref{fig1} compares the first and second order approximation of $P(W)$ with the histogram resulting from numerical simulations. Using the expressions for $A_0$ and $A_1$ in \eqref{res1av} and \eqref{res1var} gives rise to the Gaussian approximation shown by the red line. The data for the second order approximation shown in green were generated by numerically solving \eqref{eq2P} with the $\delta$-function in the initial condition \eqref{ic} replaced by a very narrow Gaussian. One clearly sees that the first order Gaussian approximation fails for most of the $W$-values. The second order result improves on the first order but is nevertheless still far from a satisfactory approximation for most of the distribution.

Let us also note, that although our perturbation expansion is complementary to the projection operator analysis of Speck and Seifert both approaches are, of course, related. In particular, conditions \eqref{normb_0} and \eqref{normb_1c_1} single the part proportional to $f_\lam(x)$ out of the solutions for $p_1(x,W,\lam)$ and $p_2(x,W,\lam)$ respectively. This is analogous to the effect of the projection operator $\hat{\Pi}_\lam$ introduced by Speck and Seifert \cite{SpSe}. 

\section{The ``simplest'' example}
A somewhat curious case is given by the so-called ``shifted parabola''
\begin{equation}\label{shpar}
 V(x,\lam)=\frac{k}{2}(x-\lam)^2
\end{equation} 
that is often considered to be the simplest example of a driven Langevin system. In fact in this case the distribution of work is Gaussian for {\em any} driving speed $\dlam=\eps$  \cite{MaJa,ZoCo}, 
\begin{equation}\label{solshpar}
 P(W,\lam)=\frac{1}{\sqrt{2\pi\si_W^2}}
             \exp{\left(-\frac{(W-\bW)^2}{2\si_W^2}\right)}
\end{equation} 
with 
\begin{equation}\label{bWshpar}
\bW(\lam)=\frac{\eps}{\mu}\lam + \frac{\eps^2}{\mu^2 k}(e^{-\mu k\lam/\eps}-1) \quad\mathrm{and}\quad
          \sigma^2_W(\lam)=\frac{2}{\beta}\bW(\lam)\; .
\end{equation} 
The perturbative treatment of this case is, however, somewhat subtle. First, we get from \eqref{defA_0} $A_0=0$ consistent with the fact that the free energy does not change with~$\lam$. Accordingly, $g_0$ does not depend on $\lam$ either (cf.~\eqref{resg_0}). The same holds true for $A_1=1/(\mu\beta)$. Consequently, the complete r.h.s.~of \eqref{resg_1} is independent of $\lam$ which implies $g_1\sim \lam$, i.e. generates yet another secular term in the perturbation expansion. A way to get rid of this extra complication is to introduce the ``super-slow'' time scale, $\tau$, via $t\to t+\eps \lam +\eps^2 \tau$  \cite{Bo}. Instead of \eqref{resg_1} we then find 
\begin{equation}\label{resg_1m}
 \pa_\lam g_1=-A_0\pa_W g_1 -A_1(\beta \pa_W g_0-\pa^2_W g_0)-\pa_\tau g_0\; .
\end{equation}  
Now we may consistently require $\pa_\lam g_1=0$. In this way we avoid an additional secular term and end up with a closed equation for $g_0$. Its solution coincides with \eqref{solshpar} with the first order result 
\begin{equation}
 \bW(\lam)=\frac{\eps}{\mu}\lam
\end{equation} 
as expected. Extending the expansion to the next order, we find $A_2=0$ as well as $C_2=0$. The latter is fine because $C_2$ describes the leading deviations from the Gaussian shape and in the present case $P(W,\lam)$ has to remain Gaussian to all orders. To understand the former we note that from \eqref{bWshpar} we have for the exact $\bW(\lam)$
\begin{equation}
 \pa_\lam \bW(\lam)=\frac{\eps}{\mu}(1-e^{-\mu k\lam/\eps})\; .
\end{equation}
As noted already in \cite{SpSe} the second term is non-perturbative and does not show up in perturbation theory at any order.


\section{Conclusion}
We have outlined a multiple time-scale perturbation expansion for the systematic determination of the work distribution in isothermal quasistatic processes. 
The expansion parameter is the ratio between the relaxation time of the system and the typical time scale of the driving. To leading order in this parameter {\em and for typical work values} the work distribution is Gaussian as expected from the central limit theorem. Higher order terms of the perturbative expansion yield systematic information about the deviations from the Gaussian form. However, irrespective of time scale separation the expansion breaks down when applied to work values from the tails of the distribution, which is again in accordance with the central limit theorem. When considering quantities that are not sensitive to the tail of the distribution \cite{Speck2} the Gaussian approximation works, of course, well. Often, however, the tails are crucial for averages relevant in stochastic thermodynamics \cite{Udorev,hugorev} and care must be exercised in using the Gaussian approximation. We finally note the pivotal role of the coupling to a heat bath for the validity of the Gaussian approximation: in non-isothermal quasi-static processes the work distribution may well deviate from a Gaussian form \cite{CrJa,LeDe,HoNiEn}. In the adiabatic expansion of an ideal gas, e.g., there are persistent correlations between different increments of the work which render the application of the central limit theorem impossible \cite{Lua,HoNiEn}.


\ack

We would like to thank Markus Niemann for a critical reading of the manuscript. Financial support from DFG under project EN 278/9-1 is gratefully acknowledged.

\end{document}